# All CVD Boron Nitride Encapsulated Graphene FETs with CMOS Compatible Metal Edge Contacts

Himadri Pandey, Mehrdad Shaygan, Simon Sawallich, Satender Kataria, Zhenxing Wang, Achim Noculak, Martin Otto, Michael Nagel, Renato Negra, Daniel Neumaier, Max C. Lemme

*Abstract*— We report on the fabrication and characterization of field effect transistors (FETs) based on chemical vapor deposited (CVD) graphene encapsulated between few layer CVD boron nitride (BN) sheets with complementary metal oxide semiconductor (CMOS) compatible nickel edge contacts. Non-contact Tera-hertz time domain spectroscopy (THz-TDS) of large-area BN/graphene/BN (BN/G/BN) stacks reveals average sheet conductivity >1 mS/sq and average mobility of 2500 cm$^2$/V·s. Improved output conductance is observed in direct current (DC) measurements under ambient conditions, indicating potential for radio-frequency (RF) applications. Moreover, we report a maximum voltage gain of 6 dB from a low frequency signal amplifier circuit. RF characterization of the GFETs yields an $f_T \cdot L_g$ product of 2.64 GHz·µm and an $f_{Max} \cdot L_g$ product of 5.88 GHz·µm. This study presents for the first time THz-TDS usage in combination with other characterization methods for device performance assessment on BN/G/BN stacks. The results serve as a step towards scalable, all CVD 2D material-based FETs for CMOS compatible future nanoelectronic circuit architectures.

*Index Terms*— Boron Nitride, chemical vapor deposition, edge contacts, graphene, mobility, radio frequency, tera-hertz, intrinsic voltage gain, voltage amplifier.

## I. INTRODUCTION

SINGLE layer graphene, first reported experimentally more than a decade ago [1], [2] has been a fast emerging electronic material with large room temperature low field carrier mobility [3]. CVD graphene is large area variant of this material, which has been shown to be useful for highly scalable devices [4]–[6]. Several applications have been suggested, demonstrated and reported in literature based on exfoliated, CVD as well as epitaxial graphene layers on various substrates in recent years. These include field effect transistors [7]–[9], inverters [10]–[14], microwave/radio frequency transistors [7], [15]–[19], various circuit applications [15], [20], [21] pressure sensors [22], [23], gas sensors [24], [25], and many more. The importance of the substrate for graphene mobility and – as a consequence – for graphene transistor performance have been outlined both through theoretical [26] as well as experimental studies [27]–[30]. Hexagonal boron nitride (hBN) has been identified by various authors as an ideal substrate for preserving nearly intrinsic, high mobility values in graphene. This has been attributed to the atomically flat surface of hBN, as well as minimal lattice mismatch with graphene [28], [31]–[35]. In addition, poor current saturation in the output characteristics (source-drain current $I_{DS}$ versus source-drain voltage $V_{DS}$) of single layer GFETs is a well-known bottleneck that results in poor voltage gain and limits $f_{Max}$ performance [36]. Good current saturation, i.e. lower $g_{ds}$, improves this situation, and is thus a desired quantity [37], [38]. For this, exfoliated hBN encapsulated monolayer graphene devices with strongly quasi-saturating output characteristics have been suggested in literature [31]. However, most of these studies have utilized mechanically exfoliated hBN crystals for device fabrication, which limits its applicability as a scalable technology. Nonetheless, there have also been some attempts to grow large area hBN sheets using chemical vapor deposition (CVD) techniques [39]. However, to the best of our knowledge, there have been no reports of CVD BN encapsulated CVD graphene (BN/G/BN) RF FETs in literature till date. In this paper, we report on the fabrication of such devices with CMOS compatible metal edge contacts to the graphene channel, and present comprehensive characterization data of the all CVD BN/G/BN FETs including DC, RF as well as THz-TDS methods. This study is also the first one to our knowledge where THz-TDS mapping has been directly used for assessing mobility in such all CVD BN/G/BN stacks. The transistors show an $I_{on}/I_{off}$ current ratio of ~6, mobility values ranging from 1000 to 5000 cm$^2$/Vs, extracted contact and sheet resistance values of ~2 kΩ·µm and ~750 Ω/sq., respectively, and quasi-saturating output characteristics. The low DC output conductance values observed in these devices are found to be lower than conventional monolayer graphene on SiO$_2$ FETs, and are comparable to artificially stacked bilayer graphene FETs, which have recently been proposed as an improvement

This work was supported by the German Research Foundation (DFG LE 2440/2-1 and LE 2440/3-1) and the EU Graphene Flagship (785219).
H. Pandey, S. Kataria and M.C. Lemme are with Chair of Electronic Devices, RWTH Aachen University, Aachen, Germany.
S. Sawallich and M. Nagel are with Protemics GmbH, Otto-Blumenthal-Str. 25, Aachen, Germany.
A. Noculak and R. Negra are with Chair of High Frequency Electronics, RWTH Aachen University, Aachen, Germany.
M. Shaygan, D. Neumaier, Z. Wang, M. Otto and M.C. Lemme are with AMO GmbH, Advanced Microelectronic Center Aachen (AMICA), Aachen, Germany (e-mail: lemme@amo.de).

over monolayer GFETs for obtaining enhanced intrinsic voltage gain performance [37]. These aspects are cumulatively exploited in demonstrating a voltage amplifier circuit application based on these GFETs which yield voltage gain of up to ~6 dB in a resistive load scheme. RF characterization of the devices yields decent values of 2.64 GHz·µm for the $f_T \cdot L_g$ product and of 5.88 GHz·µm for the $f_{Max} \cdot L_g$ product. The highest intrinsic voltage gain of an individual transistor was measured to be 7.76.

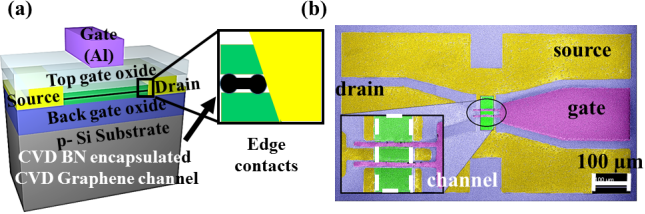

**Figure 1 (a)** Device schematic of the all-CVD BN encapsulated Graphene FET with CMOS compatible metal edge contacts. The contact metal is sputtered nickel while the top-gate metal is evaporated aluminum. **(b)** Scanning electron micrograph (color enhanced) of a device in coplanar waveguide layout, where the gate, source and drain terminals are highlighted. Inset: zoom-in view of the channel region (BN/G/BN stack). Everything except the top-gate fingers is embedded under the top-gate dielectric film.

## II. EXPERIMENT

CVD graphene monolayer (grown on Cu) was obtained from Graphenea and large area CVD grown BN was obtained from Graphene Supermarket. The materials were removed from copper using a wet etch based transfer technique and a BN/G/BN stack was fabricated by transfer onto a 1 µm thick thermally grown SiO$_2$ on p-doped Si substrate (Si resistivity ~ 1-5 Ω·cm). Optical lithography was used to pattern device channels from this material stack, using (CHF$_3$ + O$_2$) plasma based reactive ion etching. This presented us with the possibility of forming edge contacts to the encapsulated graphene sheet from the sides, as the graphene is covered under the BN sheet from the top and the bottom. Source drain contacts were then defined using optical lithography and formed using the CMOS compatible metal nickel (Ni) deposited through sputtering, followed by a lift-off process. Sputtering was preferred over evaporation because it provides isotropic metal coverage, thus facilitating the formation of edge contacts to the encapsulated graphene sheet. Ni has been demonstrated experimentally to have largest work function difference to graphene, resulting in lower graphene-metal contact resistivity [40]. Therefore, Ni was chosen for forming edge contacts to graphene in these devices. More detailed discussions on these edge contacts have been reported elsewhere [41]. The advantage of using such edge contacted graphene over regularly reported area contacted graphene is that the former shows improved contact resistivity, thus leading to improved device performance in transistor configuration [41] [42]. On top of the BN, a second top-gate dielectric of 20 nm of Al$_2$O$_3$ was deposited using atomic layer deposition (ALD) in an Oxford ALD reactor using a water vapor based cyclic process. Tri-methyl Aluminum (TMA) was used as precursor. Nucleation was facilitated thorough a 2 nm thick oxidized aluminum (Al) seed layer. Finally, top-gate fingers were fabricated using optical lithography, e-beam evaporation of 100 nm thick Al and a lift-off process in acetone. **Figure 1** (a) shows a simple schematic of such an all-CVD BN encapsulated graphene FET device. A color enhanced scanning electron micrograph (SEM) of a typical finished device in RF compatible coplanar waveguide (CPW) layout with ground-signal-ground (GSG) pads is shown in **Figure 1** (b). An atomic force micrograph (AFM) shows a stack thickness of approximately 16 nm (**Figure 2** (a)). This allows estimating the CVD BN thickness to be ~7-8 nm on each side of the monolayer graphene (~0.35 nm thick). A roughness map of the stack shows that the root mean square surface roughness was 3.1 nm (**Figure 2** (b)). On device Raman spectroscopy was used to characterize the quality of the stack to make sure that no damage occurred to the encapsulated graphene during the processing. **Figure 2** (c) shows a single Raman spectrum acquired in the channel region with clear and sharp G and 2D peaks. The monolayer nature of graphene is confirmed by a 2D/G intensity ratio > 1 [43] [6]. No significant D peak, related to presence of defects in graphene [44], was observed indicating low damage to graphene from transfer and encapsulation during the stack fabrication process. No typical hBN peak at ~1370 cm$^{-1}$ [45] could be observed in these spectra, suggesting an amorphous nature of the CVD grown BN. The uniformity of the channel regions was further confirmed by Raman area scans which yielded uniform G and 2D peak intensities over the entire channel, as shown in **Figure 2** (d).

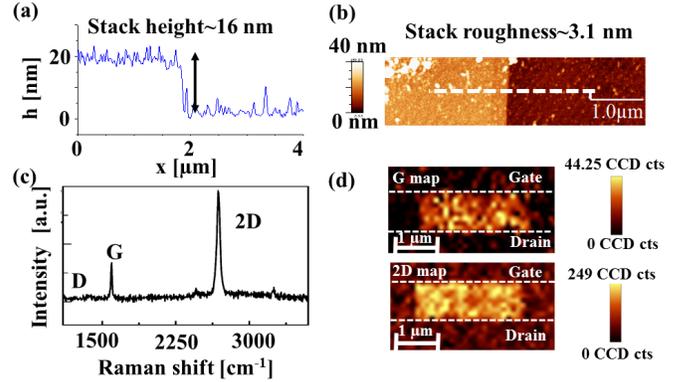

**Figure 2** (a) AFM scan of the CVD BN/G/BN stack showing a stack height of approximately ~16 nm. This indicates the presence of 7-8 nm thick BN on both sides of the monolayer graphene (thickness~0.35 nm). (b) The roughness map of the stack revealed a root mean square roughness value of 3.1 nm, measured along the white dotted line. (c) Typical Raman spectra (laser wavelength ~532 nm) of the stack. Characteristic G and 2D peaks of the encapsulated graphene monolayer are clearly seen while the D peak, which denotes defects, was found to be negligibly small. (d) Uniform G and 2D maps obtained on the sample are also shown. Both these observations suggest no damage to the graphene sheet during encapsulation between the CVD BN layers. No typical hBN peak around ~1370 cm$^{-1}$ was observed in Raman spectra, indicating a rather amorphous film.

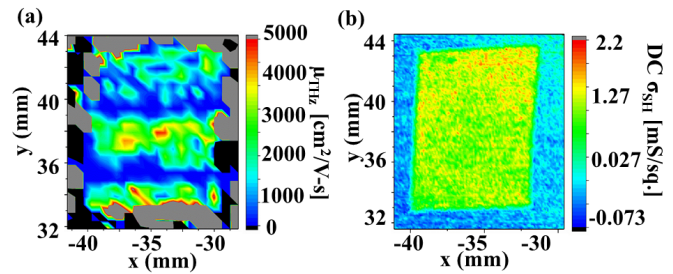

**Figure 3** (a) Spatially resolved mobility map of the all CVD BN/G/BN stack obtained from non-contact THz near-field spectroscopy indicating 1000 cm$^2$/V·s ≤ $\mu_{THz}$ ≤ 5000 cm$^2$/V·s. A median value of $\mu_{THz}$ ~ 1500 cm$^2$/V·s and an average of $\mu_{THz}$ ~2500 cm$^2$/V·s was observed over the whole stack. (b) Sheet conductivity map of the stack shows average $\sigma_{SH}$ ≥ 1 mS/sq. for the encapsulated Graphene sheet.

## III. RESULTS & DISCUSSIONS

Prior to device fabrication, the whole CVD BN/G/BN stack was characterized using a THz-TDS setup equipped with micro-probe detectors for high-resolution THz near-field imaging [46]. A THz plane wave is generated below the sample, transmitted through it and detected by a near-field detector that can be scanned at a small distance above the sample surface. **Figure 3** (a) shows the THz mobility ($\mu_{THz}$) map of the stack, with values ranging from 1000 cm$^2$/V·s to 5000 cm$^2$/V·s. It should be noted that the parameter $\mu_{THz}$ describes electron mobility values assessed using THz-TDS. A variation in THz (electron) mobility values between 1000 to 5000 cm$^2$/V.s suggested that device variability may be significant. The median of the $\mu_{THz}$ values was ~1500 cm$^2$/V·s, while the average $\mu_{THz}$ was ~2500 cm$^2$/V·s. The THz mobility map has a lateral scanning step-size of 500 μm. At each pixel, we recorded a full THz transient and obtained the spectrally resolved THz transmission amplitude of the BN/G/BN stack via Fast Fourier Transformation (FFT). From this data, the frequency dependent graphene conductivity has been calculated and the mobility has been extracted by a Drude model fit to the real part of the conductivity [47]. The details of THz-TDS based mobility extraction method for various environments have already been reported extensively in literature [47]–[49]. The large variation in $\mu_{THz}$ observed here may be attributed to variations in the fit of the measured (real) conductivity data with the Drude model. **Figure 3** (b) shows the (quasi-DC) sheet conductivity ($\sigma_{SH}$) map of the BN/G/BN stack extracted from a measurement of the THz transmission amplitude with a spatial resolution of 100 μm. The sheet conductivity and the sheet resistance ($R_{SH} = 1/\sigma_{SH}$) of the material are calculated from the ratio of the THz amplitude transmitted through the sample stack area to the substrate-only transmission amplitude using the Tinkham formula [50]. We measured quite high average values of $\sigma_{SH}$ ≥ 1 mS/sq. for the BN/G/BN stack. A corresponding $R_{SH}$ value of ~750 Ω/sq. was extracted over the scan area of the sample, as shown in **Figure 4** (a). In contrast to mobility extraction, which require full THz spectra, the sheet resistance/conductivity calculation requires only the THz amplitude. Therefore, the latter can be aquired significantly faster (or with a higher resolution in a fraction of the measurement time).

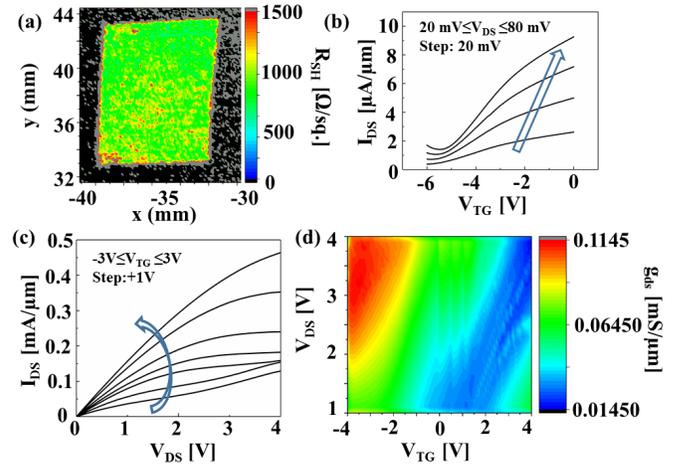

**Figure 4** (a) Extracted sheet resistance ($R_{SH}$) map of the CVD BN/G/BN stack. Average $R_{SH}$ values of ~750 Ω/sq. were observed in the sample. (b) Transfer characteristics of a BN/G/BN FET with a gate length of $L_G = 10$ μm and a channel width of W = 30 μm. The devices in general were observed to be n-doped. The arrow provides a guide to the eye in the direction of successively increasing $V_{DS}$ values. (c) Output characteristics with clear saturation of the device shown in (b). The direction of successively increasing applied $V_{TG}$ from -3V to +3V in a step of +1V is indicated by the arrow. (d) Output conductance ($g_{ds}$) map of the same device reveals improved minimum $g_{ds}$~0.01450 mS/μm, which is lower compared to non-BN encapsulated monolayer GFETs. We note that this value is comparable to minimum $g_{ds}$ values observed in artificially stacked bilayer GFETs (ASBLG FETs) [37].

All reported DC characterization was carried out under ambient conditions in a Cascade Summit 12000 probe station connected to an HP 4155B semiconductor parameter analyzer. As shown in **Figure 4** (b), transfer characteristics (source-drain current, $I_{DS}$ versus top-gate voltage, $V_{TG}$) as a function of successively increasing source-drain bias ($V_{DS}$) measured on a GFET revealed a rather highly n-doped device: the Dirac point was observed to be located at around $V_{TG}$ = -5 V. The device gate length and channel width were $L_G = 10$ μm and W = 30 μm, respectively. Such n-type doping was observed for all devices. Strongly saturating output characteristics ($I_{DS}$ versus $V_{DS}$) as a function of successively increasing applied $V_{TG}$ of the same device are shown in **Figure 4** (c). Again, similar behavior was obtained for several devices, suggesting good gate control over the channel. It should be noted that the back gate voltage was $V_{BG} = 0$V in all these data. This has been deliberately chosen to assess the RF performance potential of these devices, as will be discussed in the later part of this section. Low field carrier mobility values were extracted from transfer characteristics measured on several devices by fitting to a variable resistor GFET model reported in [51]. The best fit to the transfer curve data obtained for a device with ($L_G = 10$ μm, W = 20 μm) at $V_{DS} = 10$ mV and $V_{BG} = 0$ V resulted in maximum electron mobility of $\mu_e$ ~ 3500 cm$^2$/V·s, maximum hole mobility $\mu_h$ ~ 2500 cm$^2$/V·s and contact resistance values of ~2 kΩ·μm, respectively. The mobility values are in line with the $\mu_{THz}$ values derived from the BN/G/BN stack prior to device fabrication. The slightly lower values obtained from contact mode (field effect) mobility compared to non-contact mode mobility ($\mu_{THz}$) can be explained by considering the effects of finite contact resistance and additional fabrication process steps carried out after the THz measurements. This possibility of device variability pointed out during THz-TDS also reflected in fabricated devices, where extracted (low field, electron) mobility values were observed to vary between ~600 to 2500 cm$^2$/V.s. **Figure 4** (d) shows a detailed DC data map ($g_{ds}$ versus $V_{DS}$ and $V_{TG}$) of a GFET with $L_G$ = 10 µm and W = 30 µm, which allows choosing an optimized operating point for RF measurements. The minimum output conductance ($g_{ds}$) is ~0.01450 mS/µm. We note that this is an improvement over the minimum $g_{ds}$ values reported for monolayer GFETs fabricated on conventional SiO$_2$/Si substrates in previous work, where 0.02 mS/µm ≤ $g_{ds}$ ≤ 0.05 mS/µm [38]. The study in [38] was carried out using exfoliated monolayer graphene, which is monocrystalline, unlike CVD graphene used in this work. We also note that this value is comparable to the minimum $g_{ds}$ values observed in artificially stacked bilayer graphene (ASBLG) FETs, which have recently been proposed as an improvement over the conventional monolayer GFET for obtaining enhanced intrinsic voltage gain [37]. The highest DC transconductance measured in BN/G/BN devices was 538 µS/µm. The highest measured intrinsic voltage gain ($g_{m,max}/g_{d,min}$) in these devices was 7.76. It is important to note that the improved current saturation tendency and good voltage gain performance in the all CVD GFETs is observed despite of the presence of a large series resistance due to 3 µm access regions on both sides of the gate.

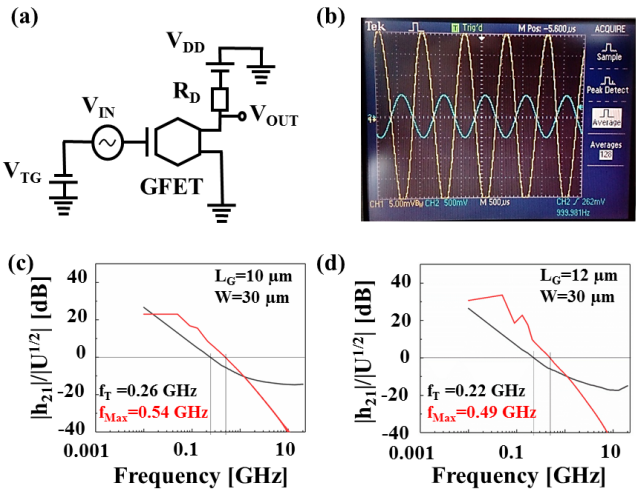

**Figure 5** (a) Schematic of a low frequency voltage signal amplifier circuit with resistive load scheme to assess the voltage gain performance of the GEFTs. A load resistance ($R_D$) of 10 kΩ was mounted on the drain end. (b) Screen shot of Oscilloscope showing the highest measured voltage gain of ~6 dB at 1 kHz for a device with $L_G$ = 10 µm and W = 10 µm.. The applied signal is shown in blue while the amplified output signal is shown in yellow on the screen. The AC input voltage was 20 m$V_{PP}$ and the measured output voltage was 40 m$V_{PP}$. (c) Highest $f_T$ ~ 0.26 GHz and $f_{Max}$ ~ 0.54 GHz values measured for the device shown in **Figure 4** with $L_G$ = 10 µm and W = 30 µm. (d) Highest $f_{Max}/f_T$ ratio of ~2.21 measured on another device with $L_G$ = 12 µm and W = 30 µm, indicating the effect of lower minimum output conductance values on RF device performance in these devices.

Furthermore, the BN/G/BN FETs were also tested for low frequency amplifier circuit applications. **Figure 5** (a) shows a simple schematic of a low frequency amplifier circuit in resistive load scheme, which was used to assess the extrinsic voltage gain performance of these devices. Several devices were measured with 1 kHz input frequency signal applied to the gate with a resistive load of 10 kΩ mounted at the drain end. **Figure 5** (b) shows a photograph of the high impedance oscilloscope used to monitor the output voltage. The highest measured voltage gain of ~6 dB was observed for a device with $L_G$ = 10 µm and W = 10 µm, when the applied AC input voltage was 20 m$V_{PP}$ while the measured output voltage was 40 m$V_{PP}$. This result demonstrates the potential capability of these all-CVD BN/G/BN FETs to be employed in circuit applications.

These devices were further tested for RF performance potential. Devices were characterized in a band width of 10 MHz to 25 GHz using calibrated Rhode & Schwarz ZVA 50 Vector Network Analyzer (VNA) connected to a compatible probe station. Calibration was performed using a standard substrate provided by the manufacturer using the Short-Open-Load-Through (SOLT) method and 2-port S-parameter measurements were performed using on-wafer GSG probes. The highest (extrinsic) maximum oscillation frequency $f_{Max}$ of ~0.54 GHz and the highest (extrinsic) cut-off frequency $f_T$ of ~0.26 GHz for a device with $L_G$ = 10 µm and W = 30 µm was observed at V$_{TG}$ = 1 V and V$_{DS}$ = 4V. (**Figure 5** (c)). The highest $f_{Max}/f_T$ ratio was ~2.21 for a device with $L_G$ = 12 µm and W = 30 µm at V$_{TG}$ = 0 V and V$_{DS}$ = 4 V (**Figure 5** (d)). The average value of $f_{Max}/f_T$ ratio in these devices was approximately 2. The state of the art $f_{Max}/f_T$ ratio of 3.2 has been demonstrated for similar micrometer sized devices with access region length of ~100 nm [52], thus minimizing the effect of parasitic access resistance. We have compared our edge contacted RF GFET devices with top-contacted RF GFET devices reported in literature for benchmarking, as it was the only available data for devices having comparable device dimensions. Nonetheless, the improved $f_{Max}/f_T$ ratio reflected in the high frequency behavior and good low frequency amplifier gain of these all CVD BN/G/BN devices are attributed to the low DC output conductance values observed in these devices (**Figure 4** (d)). The highest $f_T \cdot L_G$ product, which reflects the speed of the GFET, was 2.64 GHz·µm. The highest $f_{Max} \cdot L_g$ product observed in these devices was 5.88 GHz·µm. The RF figures of merit of these proof of concept, all CVD BN/G/BN FETs are promising given the long transistor channel lengths are used. Therefore, these devices offer good scope for future research. We also note that although the devices shown here are equipped with CMOS compatible metal (Ni) edge contacts to graphene, nevertheless, all the device fabrication steps needed for graphene FETs may or may not always be completely CMOS compatible. Some process steps used here such as chemical vapor deposition qualify for Back End of Line (BEOL) CMOS processes.

## IV. CONCLUSION

In this paper, we report large area all CVD BN/G/BN FETs with CMOS compatible metal (Ni) edge contacts. Devices have been characterized in ambient at room temperature. The device potential of such encapsulated graphene sheets has been assessed by contact (DC, RF and low frequency circuit) as well as non-contact (THz spectroscopy) methodologies. THz-TDS mapping has been used for the first time on an all CVD BN/G/BN stack. Both methodologies suggest mobility values between 1000 and 5000 $cm^2/V \cdot s$ in the BN encapsulated graphene. Good contact and sheet resistance values of around 2 k$\Omega \cdot \mu$m and ~750 $\Omega$/sq. have been extracted from these measurements. The all CVD BN/G/BN FETs yield good intrinsic voltage gain, good low frequency amplifier voltage gain as well as decent RF performance figures of merit. The complete set of DC, low frequency circuit performance as well as RF data in these devices is quite promising for future optimization. The devices demonstrate the potential for scalable, CMOS compatible all CVD BN/G/BN FETs for future device and circuit applications.